\documentclass{article}

\usepackage{amssymb}
\usepackage{graphicx}
\newcommand{\be}{\begin{equation}}
\newcommand{\ee}{\end{equation}}
\newcommand{\bea}{\begin{eqnarray}}
\newcommand{\eea}{\end{eqnarray}}

\usepackage{color}
\usepackage{hyperref}

\begin{document}

\title{Spatial Curvature in Cosmology Revisited}

\author{A. A. Coley:\\
	Department of Mathematics and Statistics, 
	Dalhousie University,\\
	Halifax, Nova Scotia, B3H 4R2, Canada}

\maketitle
	
\begin{abstract}

It is necessary  to make assumptions in order to derive models to be used for cosmological predictions and comparison with observational data. In particular, in
standard cosmology the spatial curvature is assumed to be constant and zero (or at least very small). But there is, as yet, no fully independent constraint with an appropriate
accuracy that gaurentees  a value for the magnitude of the effective normalized spatial curvature
$\Omega_{k}$  of  less than approximately $0.01$. 
Moreover, a small non-zero measurement of 
$\Omega_{k}$ at such a level perhaps indicates that the assumptions in the standard model are not
satisfied. It has also been increasingly emphasised that spatial curvature is, in general, evolving in relativistic cosmological models.
We review the current situation, and conclude that the possibility of such a non-zero
value of  $\Omega_k$  should be taken seriously.

\end{abstract}


\maketitle






\newpage
\paragraph{Introduction:}

Cosmology concerns the large scale behaviour of the Universe within a theory of gravity, 
which is usually taken to be General Relativity (GR).
The ``Cosmological Principle'', which asserts that 
{\em{on large scales the Universe can be well--modeled by a solution to Einstein's field equations (EFE) which is spatially homogeneous and isotropic}},
leads to the background Friedmann-Lema\'{i}tre-Robertson-Walker (FLRW) model
(with constant spatial curvature) with the cosmological constant,
$\Lambda$, representing 
dark energy and CDM is the acronym for cold dark matter (or  so-called $\Lambda$CDM  concordance cosmology
or {\em{standard cosmology}} for short).
Early universe inflation is often regarded as a
part of the standard model. The background spatial curvature of the universe, characterized by the normalized curvature parameter, $\Omega_{k}$, is predicted to be negligible by inflationary models
\cite{Martin19}.
{\em{Regardless of whether inflation is regarded as
part of the standard model or not, spatial curvature is assumed zero}}.

One of the greatest challenges in cosmology is understanding the origin of
the structure of the universe. 
Under the hypothesis that cosmic structure grew out of small initial fluctuations, we
can study their evolution on sufficiently large scales using {\em{linear perturbation theory}} (LPT).
The spatially inhomogeneous perturbations
exist on the uniform {\em{flat}} FLRW background spacetime.
Cosmic inflation provides a causal mechanism for primordial cosmological perturbations,
through the generation of quantum fluctuations in the inflaton field, which act as seeds for the observed anisotropies in the  cosmic microwave background (CMB)
and large scale structure  of our universe.
At late times and sufficiently small scales  (much smaller than the Hubble scale) fluctuations of the cosmic density are not small. 
LPT is then not adequate and 
clustering needs to be treated  non-linearly.
Usually this is studied with non-relativistic N-body simulations. Recently
cosmological non-linear  perturbations have been studied
at second-order and beyond \cite{MalikWandssecond}. 
More generally, non perturbative relativistic effects have been studied, especially computationally 
\cite{Adamek}
(but contrary to claims that they are ``fully relativistic'', the
assumptions in the standard model are often carried over into numerical simulations \cite{CE}).

Standard cosmology
has been very successful in describing current observations
up to various possible
anomalies \cite{tension}, which  includes the tension between 
the recent determination of the local value of the Hubble constant based on direct measurements of supernovae \cite{R18} and the value derived from the most recent  CMB 
data 
(although we note that a non-zero emergent spatial curvature may alleviate this tension  somewhat \cite{Bolejko2018,CCCS}).
In addition, since the Universe is not isotropic
or spatially homogeneous on local scales, 
the effective gravitational FE on large scales should be obtained by averaging
the EFE of GR, after which 
a smoothed out macroscopic geometry and macroscopic matter fields is obtained.
The averaging of the EFE for local inhomogeneities
can lead to significant dynamical {\em backreaction} effects
on the average evolution of the
Universe and
can affect the interpretation of cosmological data \cite{BC}. However, 
it is unlikely that  averaging  can replace dark energy, although
it can certainly affect precision cosmology at the level of 1 \% \cite{Adamek}
and may offer a better understanding of the emergence of a homogeneity scale and
of non-zero  spatial curvature \cite{CE}.

\newpage
\paragraph{Assumptions:}
It is necessary  to make assumptions to derive models to be used for cosmological predictions and comparison with observational data. But it is important to check whether the assumptions ``put in'' affect the results that ``come out''
(e.g., is the
reason that small backreaction effects are obtained in computations 
because they are assumed small from outset). In addition, we 
can only confirm the consistency of assumptions and we cannot rule out alternative explanations. 
The assumption of a FLRW background on cosmological scales presents a number of problems \cite{CE}.
For example, it is not possible to test spatial homogeneity, even in principle,
due to the existence of horizons, and we can only observe the Universe on or within the past light cone.

In particular, the assumptions that underscore the use of a 1+3 spacetime split  and a global time and a
background {\em{inertial coordinate system}}
(Gaussian normal coordinates which are  approximately Cartesian and orthogonal)
over a complete Hubble scale `background' patch in the standard model lead to the simple conditions that the spatial curvature (and the vorticity) must be very small.
The assumptions for the existence of exact 
{\em{periodic boundary conditions}} (appropriate on scales comparable to the
homogeneity scale)  imply necessarily that the
spatial curvature is exactly zero \cite{Adamek,CE}.
Any appropriate approximation will amount to  $\Omega_k$ being less than the perturbation  (e.g., LPT) scale.
It is sometimes stated that the position at which boundary conditions are assumed can be extended (above the horizon scale) in order to  avoid non-causal effects in our local patch, but this still necessitates a small spatial curvature. 
In the actual standard model the Universe is taken to be simply connected and hence the background is necessarily flat.

There are also assumptions behind the weak field approach, the applicability of
perturbation theory, Gaussian initial conditions, etc., that include
neglecting spatial curvature.
It is often claimed that backreaction can be neglected, but in LPT
the fluctuations are assumed Gaussian, which means that at the linear level all averages are zero  
by construction.
We should be cogniscent that
any intuition based on Newtonian theory may be misleading; e.g., 
the average spatial curvature of voids and clustered matter (with negative and positive curvature, respectively) is {\em{not}} necessarily zero within GR.

In standard cosmology the spatial curvature is assumed to be zero, or at least very small and  at most first order in terms of the perturbation approximation, in order for any subsequent analysis to be valid. 
Any prediction larger than this indicates an inconsistency in the approach. There is also a limit on the possible size of the observed
spatial curvature due to cosmological variance, and hence any prediction of such a small spatial curvature may not make any sense.
The standard model cannot be used to {\em{predict}}
a small spatial curvature.

\paragraph{Spatial curvature revisited:}
Current constraints on the background spatial curvature
within the standard cosmology are
often used to ``demonstrate''  that it is dynamically negligible,  $\Omega_k \sim  5\times 10^{-3}$, primarily based on CMB data. However,
the recently measured temperature and
polarization power spectra of the CMB provides a 99\% confidence level detection
of a negative 
$\Omega_{k} = - 0.044$~ $(+0.018,
-0.015$), which corresponds to a positive spatial curvature \cite{planck2018}.
Direct measurements of the spatial
curvature $\Omega_{k}$  using low-redshift data  such as supernovae, baryon acoustic
oscillations and Hubble constant observations do not place tight
constraints on the spatial curvature and allow for a large
range of possible values (but generally do include spatial flatness).
However, low-redshift observations often rely on some  CMB priors \cite{Ratra2017} and, in addition, are sensitive
to other assumptions such as the nature of dark energy.

Attempts at a consistent analysis of CMB anisotropy data in the non-flat case suggest a closed model with  $\Omega_{k} \sim 1 \%$. Including low-redshift data,
$\Omega_k =-0.086 \pm 0.078$ was obtained \cite{PR2018}, which provides weak evidence in favor of a closed spatial geometry, with stronger evidence for closed spatial hypersurfaces coming from dynamical dark energy models
\cite{XuHuang}.
As a theoretical illustration, in the phenomenological {\emph{two curvature model}}  \cite{CCCS}
(which has a parametrized backreaction contribution \cite{Coley:2005ei} leading to decoupled spatial curvature
parameters $\Omega_{k_g}, \Omega_{k_d}$ where  $\Omega_{k_g} = \Omega_{k_d}$ in the standard cosmology),  
it was found that the constraints on $\Omega_{k_g}$ and $ \Omega_{k_d}$ are
significantly weaker than in the standard model, with 
constraints on $\Omega_{k_g}$ an order of magnitude tighter than
those on $\Omega_{k_d}$ and hints from Bayesian model selection statistics that the data 
favor $\Omega_{k_d} \neq \Omega_{k_g}$.

Observations of  present-day {(large-scale)} average spatial curvature are weak and not easy to measure \cite{Larena09template}. In addition,
such measurements are affected by GR effects; e.g.,  constraints on  $\Omega_k$ may be strongly biased if cosmic magnification is not included in the analysis \cite{DiDio}. 
Given that current spatial curvature upper limits are at least one order of magnitude away from the level required to probe most of these effects (but also away from the limiting threshold) \cite{DiDio}, there is an imperative to continue attempting to constrain $\Omega_k$ to greater precision (i.e., to  a level of about 0.01) \cite{Jimenez},
especially in future surveys such as the Euclid satellite.
Most importantly, explicit model independent and CMB-independent checks of the cosmic 
flatness are necessary.

Currently there is no fully independent constraint with an appropriate
accuracy for a value of
$\Omega_{k}$  of  less than  $\sim 0.01$  from cosmological probes. 
In principle, and as noted above, a small non-zero measurement of 
$\Omega_{k}$ perhaps indicates that the assumptions in the standard model are not met,
thereby motivating models with curvature at the level of a few percent. Such models
are definitely not consistent with inflationary models
in which $\Omega_{k}$ is expected to be negligible
(and certainly below $10^{-5}$ \cite{Martin19}).
We also note that an observation of non-zero spatial curvature, even at the level of a percent or so, could be  a signal of non-trivial backreaction effects.
Analytic calculations of averaging yield a positive constant curvature 
of about this magnitude \cite{Coley:2005ei}, but numerical
estimates tend to suggest a negative mean curvature \cite{Larena09template}.

If the geometry of the universe does deviate, even slightly, 
from the standard FLRW geometry, then the spatial curvature
will no longer necessarily be constant and any effective spatial flatness may not
be preserved. A study of a {\em{small}} emerging spatial
curvature can be undertaken by relativistic cosmological simulations \cite{Adamek}.
Although such simulations need to include all relativistic corrections, they do seem to
imply deviations in the standard cosmological parameters of
a few percent on small scales below the homogeneity
scale.

However, attempts  to study relativistic inhomogeneous models that are ``close to''  FLRW {\em{can not}}  be used to address cosmological backreaction, which 
can only be present if the  structure--emerging average spatial curvature is allowed to evolve \cite{Bolejko2018}.
A dynamical coupling of matter and geometry on small scales
which allows spatial curvature to vary 
(since it does not obey a conservation law)
is a natural feature of GR, and 
schemes that suppress average curvature evolution can not describe
backreaction but only ‘cosmic variance’ \cite{BC}.
In principle,
large effects are possible from inhomogeneities and  averaging.
Recently, a fully relativistic
simulation was presented \cite{Bolejko2018} in which the perturbations are allowed to have
non-zero spatial curvature. Initially, the negative curvature
of underdense regions is compensated by the positive
curvature of overdense regions. But once the evolution enters
the non-linear regime, this symmetry is broken and the
mean spatial curvature of the universe slowly drifts from
zero towards negative curvature induced by cosmic voids
(which occupy
more volume than other regions). 
The results of this simulation indicate a
present-day value of 
$\Omega_k \sim 0.1$, as compared to the effective spatial flatness of the
early universe.

Finally,  we ask what is the current value of $\Omega_k$
necessary to question the validity and consistency of the 
standard cosmological model.
Perhaps a naive interpretation of the statistical data
implies that it is one hundred to one thousand times more likely that the value of $\Omega_k$ is in a range inconsistent with, rather than in a range  consistent with, the standard model. 
However, it is not the intent
here to suggest that the data are necessarily inconsistent with the standard model, but rather to 
assert that the possibility of  a significant non-zero
value of  $\Omega_k$  should be taken seriously.

\newpage

\paragraph{Acknowledgements:}   Financial support was provided by  NSERC of Canada.

\end{document}